# Electron sampling depth and saturation effects in perovskite films investigated by soft x-ray absorption spectroscopy

A. Ruosi[1,*], C. Raisch[2], A. Verna[3], R. Werner[4], B. A. Davidson[3], J. Fujii[3], R. Kleiner[4] and D. Koelle[4]

[1] Dept. of Physics, University of Naples "Federico II", P. Tecchio 80, 80125 Naples, Italy

[2] Physikalische Chemie and LISA$^+$, Universitat Tuebingen, 72076 Tuebingen, Germany

[3] CNR-IOM TASC National Laboratory, AREA Science Park, 34012 Basovizza, Trieste, Italy

[4] Physikalisches Institut and Center for Collective Quantum Phenomena in LISA$^+$, Universitat Tuebingen, 72076 Tuebingen, Germany

Knowledge of the electron sampling depth and related saturation effects is important for quantitative analysis of X-ray absorption spectroscopy data, yet for oxides with the perovskite structure no quantitative values are so far available. Here we study absorption saturation in films of two of the most-studied perovskites, $La_{0.7}Ca_{0.3}MnO_3$ (LCMO) and $YBa_2Cu_3O_7$ (YBCO), at the $L_{2,3}$ edge of Mn and Cu, respectively. By measuring the electron-yield intensity as a function of photon incidence angle and film thickness, the sampling depth $d$, photon attenuation length $\lambda$ and the ratio $\lambda/d$ have been independently determined between 50 and 300 K. The extracted sampling depth $d_{LCMO} \approx 3$ nm for LCMO at high temperatures in its polaronic insulator state (150 – 300 K) is not much larger than values reported for pure transition metals ($d_{Co\ or\ Ni} \approx 2 - 2.5$ nm) at room temperature, but is smaller than $d_{YBCO} \approx 3.9$ nm for metallic YBCO that is in turn smaller than the value reported for $Fe_3O_4$ ($d_{Fe3O4} \approx 4.5$ nm). The measured $d_{LCMO}$ increases to 4.5 nm when LCMO is in the metallic state at low temperatures. These results indicate that a universal rule of thumb for the sampling depth in oxides cannot be assumed, and that it can be measurably influenced by electronic phase transitions that derive from strong correlations.

I.  INTRODUCTION

X–ray absorption spectroscopy (XAS) in solids involves the excitation by photon absorption of a core level electron into unoccupied states near or above the Fermi level. The subsequent filling of the core hole by an electron with lower binding energy results in either the emission of a fluorescence photon or the radiationless emission of an Auger electron. Auger decay largely dominates over fluorescence decay for all core levels below 1 keV of binding energy [1, 2]. The depth sensitivity of the XAS technique depends on which decay products are collected, photons ("fluorescence yield" or FY) or electrons ("electron yield" or EY). The much smaller interaction cross-section of x-rays in solids, as compared to electrons, makes the EY method much more surface sensitive than the FY method; the sampling depth $d$ in EY mode is material dependent and typically <7 nm, while it is of the order of 100 nm for FY mode for photon energies <1 keV. Electron yield methods can be further refined into the partial electron yield (PEY) method that collects escaped electrons only in a selected energy window, while the total electron yield (TEY) method detects all primary photoelectrons, Auger and secondary electrons that emerge from the sample surface independent of their energy. In this way, the TEY method takes advantage of the avalanche of low kinetic energy secondary electrons produced through inelastic electron-electron scattering from the primary high-energy Auger photoelectrons that leads to a larger measured signal.

Knowledge of the electron sampling depth $d$ is fundamental to extract information about a buried layer or interface from TEY spectra, but its prediction is a cumbersome task. Indeed, $d$ depends not only on photon energy but also on atomic species, composition and the material's crystallographic and electronic structure; its magnitude depends on the electron scattering efficiency inside the investigated material (see for example [3]). Thus, theoretical treatments should include interactions of direct and indirect secondary electrons with that particular solid through all elastic and inelastic scattering channels. Models describing these decay processes as a function of electron energy have not yet been successful in predicting sampling depths in many materials, despite the fact that TEY absorption spectroscopy has been used intensively for years to investigate surfaces and interfaces in a wide range of materials. As a consequence, except for the few materials for which the sampling depth has been experimentally determined, the depth sensitivity of this technique is still under discussion. Previous studies on transition metals and rare earths showed a strong dependence of $d$ on the investigated material, in the range 1.5 – 2.5 nm for metallic Fe, Ni, Co, Dy and Tb for photon energies in the range 700 – 900 eV [4, 5, 6, 7, 8, 9]. Because of the different electronic structure between metals and insulators, in particular the gap in the density of states near the Fermi level in the latter that will influence the secondary electron cascade, insulators are expected to have a larger electron sampling depth. Little literature is available on this topic [9, 10, 11, 12]. Gota et al. [11] found an unexpectedly high value for the sampling depth of the iron magnetic oxide $Fe_3O_4$ of 4.5 nm, in comparison to the one known for metallic Fe in the range 1.7 – 2.2 nm, while $d \approx 2$nm was found both in $Ta_2O_5$ at the O $K$- edge [9] and in LaFeO$_3$ at the Fe $L_{2,3}$ edge [12]. Recently, transition-metal oxides with the perovskite structure and their interfaces have been intensely investigated, especially using XAS techniques to probe their interfaces [13, 14] because of their potential application in spintronic or superconducting devices. Surprisingly, experimental determination of electron sampling depths for many of these technologically-relevant oxides has yet to be performed.

Measurements of the polarization-dependent photoabsorption cross-section in x–ray linear and magnetic circular dichroism techniques (XLD and XMCD, respectively) have enabled the extraction of element-specific expectation values such as orbital occupations, number of holes in the *d*-bands, and spin and orbital magnetic moments from heterostructures [15, 16, 17]. To obtain reliable quantitative results from the observed electron yield signal, investigation of the extent of the saturation effect in the absorption process is essential [6, 7, 11, 18, 19, 20, 21, 22, 23]. In TEY, saturation arises when the electron sampling depth $d$ becomes comparable to the photon penetration depth $\lambda \sin\theta$, i.e. the projection of the attenuation length $\lambda$ along the surface normal ($\theta$ is the grazing angle from the sample surface): as the ratio $d/\lambda\sin\theta$ increases, saturation effects become more evident. Then, in presence of a large electron sampling depth or alternatively when either a strong attenuation length or a sufficiently small incidence angle (θ < 30°) is used, corrections using an empirical yield equation have to be performed on the measured spectra [6, 7, 23]. Indeed, in this case the spectral shape of the signal can significantly deviate from the absorption coefficient since the assumption of proportionality between electron yield and absorption coefficient is not longer fulfilled in the energy range of interest. The overall XAS amplitude decreases and, in particular, the high intensity peaks will be more reduced in comparison to the low intensity ones; also, the sampling depth will then appear smaller. In sufficiently thin films a small saturation can affect absorption spectra even at normal photon incidence (~3% effect in a 3 nm film is reported in Ref. [7]). Modeling of absorption saturation on Fe, Co and Ni $L_{2,3}$-edge spectra recorded by electron yield detection demonstrates that the error bars introduced by this spectral distortion can be in excess of 100% on the orbital magnetic moment and 10-20% on the number of *d*-holes and the spin moment [17]. The occurrence of angular-dependent saturation can be a limitation in polarization–dependent XAS experiments in which low grazing angles (10°–30°) must be chosen. Moreover, it has been suggested that saturation effects may be generally stronger for oxides than for the corresponding transition metal because of the larger sampling depth found in Fe$_3$O$_4$ [11]. Removal of saturation effects is a delicate task since it requires inversion of the electron yield function that has a nonlinear dependence on energy. From a practical point of view, at a fixed angle the different yield reductions due to saturation at different resonances makes it difficult to extract quantitative results from absorption spectra unless the sampling depth and attenuation length at each resonance is known.

In this paper we investigate absorption saturation in perovskite films at the transition-metal $L_{2,3}$ edge, measuring Mn (~ 640 eV) in LCMO and Cu (~ 930 eV) in YBCO in well-characterized films on SrTiO$_3$ (STO) substrates. Following the approach of Ref. [11], we have measured the TEY intensity as a function of photon incidence angle and film thickness, and independently extracted the attenuation length $\lambda$, the sampling depth $d$ and their ratio $\lambda/d$. These measurements were performed at 50 – 300 K, a temperature range over which the LCMO and YBCO resistivities change by many orders of magnitude: LCMO crosses a metal–insulator transition temperature $T_{MI}$ ~ 150 K, and YBCO crosses a superconducting transition temperature $T_c$ ~ 90 K. The extracted sampling depth $d_{LCMO} \approx 3$ nm in LCMO at high temperatures (150 – 300 K) when it is in a polaronic insulating phase [24] is not much larger than values reported for pure transition metals ($d_{Co\ or\ Ni} \approx 2.5$ nm), but is smaller than $d_{YBCO} \approx 3.9$ nm for metallic YBCO that is in turn smaller than the value reported for Fe$_3$O$_4$ ($d_{Fe3O4} \approx 4.5$ nm). Furthermore, the sampling depth increases significantly in metallic LCMO below $T_{MI}$, while it remains more or less constant in YBCO below $T_c$. These results indicate that a universal rule of thumb for sampling depths in oxides cannot be assumed, that the sampling depth can be measurably influenced by electronic phase transitions due to strong correlations, and that the sampling depth for each material should be measured individually until a sufficiently accurate model is developed for its prediction. These results should encourage an investigation of sampling depths in other correlated materials, in order to enable proper interpretation of XAS spectra used to probe the orbital, charge and magnetic order in films and buried interfaces.

II. EXPERIMENTAL DETAILS

Epitaxial LCMO and YBCO thin films were grown on (001) SrTiO$_3$ (STO) substrates by pulsed laser deposition (PLD) [14]. A commercially available LCMO target was ablated by a KrF laser ($\lambda = 248$nm) with a repetition rate of 2 Hz. The substrate temperature during deposition was 750ºC and the oxygen pressure during growth was set to 20 Pa. After deposition, the chamber was flooded with oxygen to 1 mbar and the samples were cooled down at 10º/min to 550ºC, where they were annealed for 1 hour. *In situ* high-pressure reflection high-energy electron diffraction (RHEED) was used to control the exact number of deposited monolayers. The thin film resistance was measured by the Van der Pauw method in the temperature range 10 – 300 K. The 200 nm thick YBCO (100) film was grown on a 30 nm $PrBa_2Cu_3O_7$ (PBCO) template deposited on STO.

The experiments were carried out at Advanced Photoelectric Effect (APE) beamline located at the Elettra storage ring in Trieste that exploits photons in the range 140 – 1500 eV. Details of the APE beamline have been published elsewhere [25]. XAS measurements were performed in total electron yield (TEY) in current mode, on the manganese (LCMO), titanium (STO) and copper (YBCO) $L_{2,3}$ edges. An energy resolution in the energy range of interest (400 – 900 eV) better than 200 meV was achieved by setting the exit slits of the grating monochromator in the range 20 – 50 μm. The TEY signal was monitored by the drain current through the sample and normalized to the incident photon flux measured by a ~80% transmission gold mesh. The photocurrent flowing to the sample is measured by a picoammeter and recorded as a function of photon energy. Typical sample and mesh current are in the range of a few nA and 500 pA, respectively. The sample motion has four degrees of freedom, motorized $x$, $y$ and $z$ and manually operated $\theta$ (polar angle). The polar angle is determined by the direction of light incidence vector and the sample surface normal; at $\theta = 90°$ the photon wavevector is (anti)parallel to the surface normal, and at $\theta = 0°$ it is parallel to the sample surface. For the polar angle dependent yield measurements, a series of incidence angles ($\theta = 5° – 90°$ in 5° increments) was chosen. The error in the angle alignment is ~0.5°. Samples can be cooled down to 45 K via a liquid-helium flow cryostat. The angular dependence of the absorption intensity was measured at temperatures of 50 K, 150 K and 300 K, corresponding to metallic, maximum resistivity, and insulating states of LCMO, respectively, and the superconducting (50 K) and metallic states (150 K and 300 K) of YBCO. In order to consider film thickness variations, three different sets of beamline throughput curves in different points of the sample surface have been measured for each sample.

III. THIN FILM CHARACTERIZATION

Surface flatness, a precisely known film thickness and lack of film/substrate interdiffusion are important for this type of study. The first two of these were monitored *in situ*, as the number of deposited unit cells was tracked carefully during deposition and surface morphology was checked by RHEED and atomic force microscopy (AFM). RHEED results are shown in Fig. 1 for the 13 uc LCMO sample. The observed oscillations of the specular RHEED intensity are a clear indication of a layer-by-layer growth and allow the determination of the film thickness, since each maximum represents one deposited unit cell of LCMO. Fig. 2 shows thickness fringes near the (002) reflection for thicknesses of 25 uc and above, indicative of a highly uniform LCMO thickness. These measurements were used to crosscheck the thickness evaluated from RHEED intensity oscillations, and they were found to coincide within one unit cell. Film thicknesses were 7, 13, 25 and 50 unit cells (uc), or 2.7, 5.0, 9.6 and 19.2 nm, respectively, and for these thicknesses the films are fully strained with the STO substrate, with in-plane lattice parameters $a = b = 0.3905$ nm and c = 0.384 nm. The surface roughness of the LCMO samples were checked by AFM and were found to reproduce the terrace steps of the prepared STO substrates, giving an RMS roughness <0.2 nm for all thicknesses. These values were in agreement with sharp RHEED specular spot whose intensity was comparable between the starting STO substrate and the LCMO at the end of growth. In order to obtain a reliable value for XAS sampling depth $d$, nonuniform film thickness or interdiffusion with the substrate should be carefully checked and excluded, as they would tend to increase the measured value for $d$. We can estimate an upper limit for the interdiffusion between the STO substrate and the LCMO overlayer to about one unit cell [26]. Summing this to an error in thickness of ±0.5 uc gives the error bars used during fits of the sampling depth discussed in Sec. IV.1.

The resistivity versus temperature $R(T)$ for the strained 50 uc LCMO film is shown in Fig. 3. The metal-to-insulator transition temperature $T_{MI}$ is defined as the temperature for which $R(T)$ shows a maximum, typically within a few Kelvin of the ferromagnetic transition temperature $T_{Curie} \approx 160$ K in this film. This reduction in $T_{Curie}$ from the bulk value ~ 270 K is primarily due to strain and has been reported previously [27]. The LCMO resistivity changes by two orders of magnitude as the temperature is reduced from 160 K to 50 K. We note that, for the analysis of the temperature dependence of the sampling depth in Sec. IV, XAS spectra were taken at temperatures in the insulating temperature region (300 K), near the resistivity maximum (~150 K) and well into the metallic region (50 K). The YBCO film had a superconducting transition temperature $T_c$ of 88 K and was metallic above $T_c$, with a flat surface as indicated by sharp 2-dimensional RHEED features at the end of growth.

IV. RESULTS AND DISCUSSION

This section is organized as follows. For measurements on LCMO films, all three of the relevant parameters (sampling depth $d$, attenuation length $\lambda$, and their ratio $\lambda/d$) were independently determined at 300 K. Since the ratio of the measured $\lambda$ to the measured $d$ was in good agreement with the measured $\lambda/d$, at lower temperatures (150 K and 50 K) only $\lambda$ and $\lambda/d$ were independently measured and $d$ was inferred by dividing the former by the latter. Subsequently, for measurements on YBCO films, $\lambda$ and $\lambda/d$ were independently measured and $d$ was inferred by dividing the former from the latter. In Sec. IV.1 we present the direct measurements on the sampling depth $d$ in LCMO; in Sec. IV.2 we present measurements on the attenuation length $\lambda$; in Sec. IV.3 we present measurements on the ratio $\lambda/d$ and the inferred values for $d$ using $\lambda$ from the previous section. All results for $d$, $\lambda$ and $\lambda/d$ for both LCMO and YBCO films are summarized in Table 1 in Sec. IV.3. In Sec. IV.4 we discuss these results.

IV.1 SAMPLING DEPTH $d$

The experimental sampling depth $d$ quantifies the depth sensitivity of the signal measured in a total electron yield experiment. The sampling depth is determined by two basic processes: the first that governs the generation of primary and secondary electrons following the core-level absorption event, and the second that controls which electrons will ultimately escape from the sample surface. The former is limited in distance by the total path length traveled by the primary Auger electron in the solid before it has dissipated all its kinetic energy (effective penetration range $R_{Auger}$), while the latter is characterized by the electron escape depth $\lambda_{escape}$ (see discussion in Appendix). In a simple model, the secondary electrons are created within a sphere with a radius $R_{Auger}$ having the absorption site placed at the center. Their production occurs particularly at the end of the range of primary electrons, when their initial kinetic energy has fallen below 40 eV. For soft x-rays, the TEY signal corresponds to the integration of a photoelectron spectrum from the high-energy valence band photoelectrons to low-energy secondary electrons produced in the cascade, that have a broad kinetic energy distribution typically peaked below ~5 eV (see discussion in Refs. [1, 2, 9] and the Appendix). Due to the shape of the universal electron escape depth curve in which $\lambda_{escape}$ has a minimum near 1 keV and increases exponentially going towards lower energy, the TEY signal is dominated by the low-energy electrons in the cascade. Since the precise energy distribution of the cascade will depend on the atomic species, composition and structure of the material that determine $R_{Auger}$ and $\lambda_{escape}$, the resulting sampling depth is consequently difficult to predict numerically.

The sampling depth $d$ can be directly determined by measuring the electron yield in normal photon incidence for a series of films having different thicknesses $t \leq 5d$. In Fig. 4 we show the edge–jump amplitude of absorption spectra at the Mn $L_{2,3}$ edges with normal incidence plotted as a function of LCMO film thickness measured at room temperature. A linear slope that fitted the pre-edge background has been subtracted to the spectra to determine the peak height. A function that describes the emission depth profile may be derived considering the absorbing material as a stacking of adjacent infinitesimal slices each contributing to the total electron yield signal [28]. Slices that are deeper will contribution less because of scattering events that occur as electrons migrate towards the surface. Under the assumption that the number of produced electrons follows an exponential law with a decay constant $d$, the contribution to the signal of an infinitesimal slice at depth $z$ is given by $dI(z) = A\exp(-z/d)dz$, where $A$ is a constant depending on the material under examination. Then the total electron yield signal for a film having a finite thickness $t$ will be $I(z_0) = A(1-\exp(t/d))$. The exponential best-fit with $d$ as a free parameter results in a value of the sampling depth $d_{L3} = 2.6 \pm 0.3$ nm for the $L_3$ and $d_{L2} = 2.7 \pm 0.3$ nm for the $L_2$ absorption edges. The Mn signal rapidly increases for film thickness up to about 5 nm, at which thickness ~90% of the saturation value has been reached (Fig. 4); above 10 nm the curve is very nearly constant.

IV.2 ATTENUATION LENGTH $\lambda$

The attenuation length, $\lambda(E)$, defined as the depth into the material at which the x-ray intensity has decreased to 1/e (~ 37%) of its value at the surface, is another fundamental parameter in the soft-x-ray absorption process. The attenuation length is found by inverting the absolute (linear) absorption coefficient $\mu(E) = 1/\lambda(E)$ that is found by scaling the measured XAS spectra to the value obtained far from resonances (where the absorption coefficient of a given element is almost independent on the chemistry of nearby atoms and saturation effects are negligible) from tabulated atomic photoabsorbtion cross-sections $\sigma_a(E)$ [29] (see discussion in the Appendix). The experimental Mn $L_{2,3}$ spectra measured at 300 K, 150 K and 50 K on a ~20 nm thick LCMO film, scaled to the tabulated cross section (dotted lines) are shown in Fig. 5. To minimize saturation effects, normal photon incidence was chosen. The attenuation length in this energy range is determined from $\mu_a^{LCMO}(E)$. $\lambda(E)$ varies strongly across the resonance energies, and takes the value $\lambda(L_3) = 50.3 \pm 4$ nm and $\lambda(L_2) = 78.7 \pm 6$ nm at the maxima of the $L_3$ and $L_2$ edges, respectively. The attenuation length measured at the Mn $L_3$ and $L_2$ resonances in LCMO here are larger than those reported for the Fe $L_3$ and $L_2$ resonances in $Fe_3O_4$, 17 nm and 52 nm, respectively [11] and in Fe, Co or Ni, of ~20 nm and ~35 nm, respectively [6, 7, 15]; this may be expected due to the large fraction of oxygen in the perovskite lattice. As expected, within the errors, the attenuation length is temperature independent. The corresponding absorption coefficients in LCMO are $\mu(L_3) = 19.9\ \mu m^{-1}$ and $\mu(L_2) = 12.7\ \mu m^{-1}$.

The attenuation length $\lambda$ in YBCO films at the $L_3$ (931 eV) and $L_2$ (952 eV) absorption edges has been determined from the XAS spectra at the Cu $L_{2,3}$ resonances by the same method, giving values for $\lambda(L_3) = 75 \pm 6$ nm and $\lambda(L_2) = 138 \pm 9$ nm with corresponding absorption coefficients $\mu(L_3) = 13.3\ \mu m^{-1}$ and $\mu(L_2) = 7.3\ \mu m^{-1}$.

IV.3 SATURATION AND RATIO $\lambda/d$

A determination of $\lambda/d$ can be obtained from the variation of the $L_2$ and $L_3$ intensities on incidence angle, since saturation effects depend only on the ratio of $\lambda$ and $d$ and not their absolute values. Saturation occurs when the incidence angle between the light and the sample surface is sufficiently less than 90° so that the projection of $\lambda$ along the surface normal $\lambda \sin\theta$ becomes comparable to $d$. Saturation effects can be particularly relevant for materials containing transition metal and rare earth atoms with a considerable amount of unoccupied states in $d$- and $f$-shells, and consequently high absorption cross sections, and can be important at low grazing angles ($\theta \leq 10°$-$30°$) that are often necessary in some polarization-dependent experiments. Generally, the total electron yield is assumed to be proportional to the linear absorption coefficient $\mu$, though in the energy range 50–1500 eV this is strictly true only in the limit $\lambda \sin\theta/d \gg 1$ [30]. Saturation will generally reduce the TEY through the following relation between the net total electron yield of a semi-infinitely thick sample and the absorption length $\lambda$ [6]:

$I_{TEY}(E,\theta) = Ad/(d + \lambda(E)\sin\theta)$ (1)

Here the parameter $A$ is the number of cascading electrons produced per photon that migrate towards the free surface. Only for weak absorption and/or for incidence angle close to the sample surface normal, in the limit $d \ll \lambda \sin\theta$, the intensity is directly proportional to $\mu$: $I_{TEY} \approx Ad/\lambda(E)\sin\theta \approx 1/\lambda =$ $\mu(E)$. Conversely, when $d \gg \lambda\sin\theta$ there is complete saturation ($I_{TEY}$ = A) and the proportionality between $I_{TEY}$ and $\mu$ breaks down. In the range where most resonant absorption experiments are performed, typically $1 < (\lambda\sin\theta)/d < 100$, the measured signal is neither proportional to $\mu$ nor fully saturated [17]. When changing from normal to grazing incidence geometry, the electron yield signal of Eq. (1) will increase since the effective photon path length $\lambda\sin\theta$ in the active thickness $d$ decreases. Consequently the photons are absorbed in shallower layers with respect to the surface where it is easier to escape from the sample. A convenient way for verifying the presence of saturation effects is to multiply Eq. (1) by $\sin\theta$:

$$I_{TEY}(\theta)\sin\theta = A/[(\lambda(E)/d) + \csc\theta] \qquad (2)$$

that represents the total electron yield intensity normalized to the number of absorbing atoms. When $\lambda/d \gg 1$, saturation does not affect the electron yield signal and the quantity in Eq. (2) will be independent of the incidence angle.

To investigate the extent of saturation effect in manganites, we measured $I_{TEY}(E)$ at the Mn $L_{2,3}$ edge for a series of incidence angles between 5° – 90° for the 50 uc thick LCMO film at three different temperatures, 300 K, 150 K and 50 K. The $I_{TEY}(E)$ spectra are shown in Fig. 6(a). This film thickness is a good approximation of a semi-infinite sample for the TEY technique, since the XAS signal at the Ti $L_{2,3}$ resonance (460 – 490 eV) from the underlying substrate is below the signal-to-noise level of our measurement setup at all incidence angles. Apart from a normalization of the intensity $I_{TEY}(E)$ to the incoming radiation intensity (via the mesh current), no procedure of background removal has been applied. Indeed the number of secondary electrons in the background holds information on saturation effects (i.e., the pre-edge and post-edge intensities follow a saturation-like curve as a function of incidence angle). Angular dependent saturation effects are visible to the eye in the spectra as a change in the pre- and post-edge intensities, as well as the relative $L_3$ and $L_2$ peak heights, going to lower incidence angle.

The maximum $I_{TEY}$ at the $L_3$ and $L_2$ peaks are multiplied by $\sin\theta$ and plotted in Fig. 6(b) for all three temperatures. Fits to the data were performed using Eq. (2) with two free parameters, $A$ and $\lambda/d$, resulting in the same ratio $\lambda/d(L_3)$ = 15.4 ± 0.7 and $\lambda/d(L_2)$ = 23.7 ± 1.2 for the two curves measured at 300 K and 150 K. Surprisingly, the curve measured at 50 K shows a different angular dependence with respect to the curves measured at higher temperatures. Fitting the data at 50 K gives decreased values of $\lambda/d$ at both absorption resonances $\lambda/d(L_3)$ = 11.1 and $\lambda/d(L_2)$ = 17.0. Since the attenuation length is independent of $T$ (Sec. IV.2), these measurements imply a temperature-dependent sampling depth.

Saturation is more pronounced at the $L_3$ resonance than at $L_2$, since at $L_2$ the condition $\lambda(E)\sin\theta \gg d$ is better satisfied. Even at $L_2$, where $\lambda$ is more than 20 times larger than $d$, the proportionality between $I_{TEY}$ and $\mu$ is not completely fulfilled. The spectra in Fig. 6(b) show that at $\theta = 5°$, $I_{TEY}\sin\theta$ is reduced by about 40% and 25% for $L_3$ and $L_2$, respectively. Consequently, soft X-ray polarized absorption experiments of LCMO must be performed at incidence angles $\theta \geq 30°$, above which the electron yield intensity reduction at $L_3$ and $L_2$ is less than 6% and 2%, respectively.

Saturation effects in YBCO similar to those seen in LCMO are evident in the absorption spectra versus angle shown in Fig. 7(a). Values for $\lambda/d$ in YBCO at 150 K and 50 K were extracted from fits to $I_{TEY}\sin\theta$ versus incidence angle in the same way as described above and are shown in Fig. 7(b). The fits for $\lambda/d$ at 150 K yield values for $\lambda/d(L_3)$ = 19.1 ± 1.0 and $\lambda/d(L_2)$ = 25.2 ± 1.2, and are listed in Table 1. At 50 K, below the superconducting transition temperature, values for $\lambda/d$ are similar to those at 150 K.

The value of sampling depth $d$ can also be inferred from independently measured values for $\lambda$ and $\lambda/d$. For LCMO, at 300 K the inferred values are $d_{L3}$ = 3.2 ± 0.4 nm and $d_{L2}$ = 3.3 ± 0.4 nm, close to but slightly larger than the directly measured value $d_{L3}$ = 2.6 nm and at the limit of the error bars. The inferred sampling depth at both resonances is unchanged between 300 K and 150 K but increases appreciably at 50 K, accompanied by an increase in the maximum TEY intensity in the absorption spectra at low temperature and a change in curvature (Fig. 6(b)). The change in curvature leads to a fit value of $\lambda/d$ that is smaller, and consequently $d^{50K}$ is larger than its value at higher temperatures. Any temperature dependence of $d$ is unexpected, or at least has not yet been treated in the literature, and is discussed below in Sec. IV.4. Inferred values for the probing depth in YBCO are $d^{inferred} \approx 3.9 ± 0.5$ nm at Cu $L_3$ and 5.5 ± 0.6 nm at $L_2$. Values for $d$, $\lambda$ and $\lambda/d$ are summarized in Table 1 for Mn and Cu at $L_3$ and $L_2$ resonances at all temperatures measured.

| Temperature (K) | Resonance | $d$ (nm) | $\lambda$ (nm) | $\lambda/d$ | $d^{inferred}$ (nm) |
|---|---|---|---|---|---|
| 300 | Mn $L_3$ | 2.6 ± 0.3 | 50.3 ± 4 | 15.4 ± 0.7 | 3.2 ± 0.4 |
| 150 | " | — | " | 15.5 | 3.2 |
| 50 | " | — | " | 11.1 | 4.5 |
| 300 | Mn $L_2$ | 2.7 ± 0.3 | 78.7 ± 6 | 23.7 ± 1.2 | 3.3 ± 0.4 |
| 150 | " | — | " | 24.0 | 3.3 |
| 50 | " | — | " | 17.0 | 4.6 |
| 150 | Cu $L_3$ | — | 75.0 ± 6 | 19.1 ± 1.0 | 3.9 ± 0.5 |
| 50 | " | — | " | 21.3 | 3.5 |
| 150 | Cu $L_2$ | — | 138 ± 9 | 25.2 ± 1.2 | 5.5 ± 0.6 |
| 50 | " | — | " | 28.3 | 4.9 |

Table 1. Values of sampling depth $d$, attenuation length $\lambda$, and the ratio $\lambda/d$, each determined independently from the XAS spectra, or inferred from the measured $\lambda \div (\lambda/d) = d^{inferred}$, at different temperatures and different absorption resonances in LCMO or YBCO. Errors on $d^{inferred}$ are determined by the propagation of errors from $\lambda$ and $\lambda/d$.

IV.4 DISCUSSION

Using the general expression of the analytical model for the total-electron-yield depth dependence [31], we can calculate the contribution to the secondary-electron-yield current per Auger electron created at depth $\lambda_{escape}$. Even though the application of this model is correct only in the photon energy range 1–10 keV, the values of the parameters $\lambda_{escape}$ and $R_{Auger}$ that result in a sampling depth of 3 nm are reasonable: ~5 Å and ~60 Å, respectively. Typically, in metals the energy-loss mechanism for internal secondary electrons is related to interactions with conduction electrons, lattice vibrations and defects. In order to leave the sample surface, the kinetic energy of a secondary electron, calculated with respect the Fermi level, must be larger than the work function $W$ (~5 – 10 eV). The high collision probability due to the large number of conduction electrons along with the large value of minumum escape energy, results in a small sampling depth and secondary yield. Experimentally determined values for $d$ of transition metals at the $L_{2,3}$ edge including Cu are in the range 1.7 – 2.7 nm [4-9].

In band insulators, secondary electrons loose their energy through excitation of valence electrons into the conduction band. The minimum kinetic energy the electrons have to overcome to escape is the electron affinity $\chi$, i.e. the difference between the vacuum level and the conduction-band minimum ($\chi \sim 1$ eV). In this case the wide band gap inhibits secondary electrons at the bottom of the conduction band (with kinetic energy $< E_{\text{gap}}$) from participating in electron-electron collisions. Therefore the sampling depth and the electron yields are generally larger than in metals. The few experimental reports reported on insulating materials and oxides give values of $d = 2$ nm in $Ta_2O_5$ at the O $K$-edge (530 eV) [9], $d < 3.9$ nm in $Al_2O_3$ at the Al $K$-edge (1560 eV) [10], $d = 4.5$ nm for $Fe_3O_4$ [11] and $d = 2$ nm in $LaFeO_3$ [12], the latter two values at the Fe $L$-edge (710 eV). The behavior of Mott-insulators, in which electron correlations drive localization and the effective band gap can be quite small, is more cumbersome than typical oxides with large band gaps. The value we report here for $d$ measured at room temperature in LCMO when LCMO behaves as an polaronic insulator falls between the values for metals and insulators cited above. However, at low temperatures when LCMO is metallic, the probing depth unexpectedly increases. At the same time, YBCO at 150 K has roughly the same metallic resistivity of LCMO at low temperatures, and its sampling depth is larger than insulating LCMO but smaller than metallic LMCO and $Fe_3O_4$. These results imply that the idea of a larger sampling depth for oxides as compared to those of pure transition metals is not always true, in any case the differences are not as dramatic as previously believed, and each material must be looked at individually. In addition, and perhaps surprisingly, our results suggest oxides with a correlation-driven metal-insulator transition can have significantly larger probing depths when metallic than when insulating. Thus, as a practical rule, saturation effects will impact TEY measurements of manganite oxides for incidence angles $\leq 30º$, and similarly for YBCO for incidence $\leq 35º$, though somewhat less than seen in e.g. $Fe_3O_4$.

Lastly we discuss the possibility of a temperature–dependent sampling depth in LCMO inferred from the fits of $\lambda/d$ (Sec. IV.3, Table 1). It is possible that the large changes in electronic structure of LCMO across $T_{MI}$ [32] could change the efficiency of electron scattering processes inside the material at low temperatures. Any increase of the inelastic mean free path of secondary electrons would give rise to an increased electron yield with an enhanced statistical weight of lower-energetic secondary electrons migrating towards the surface. Another explanation may be related to a change in the conductivity and condition of the sample surface, resulting in a subsequent decrease of the surface work function with temperature. Typically, the majority of secondary electrons that escape from the surface have a kinetic energy in the range 3 – 5 eV [31]. Subsequently, a small reduction of the vacuum barrier at the surface could have a strong effect on the total yield. This change in work function could be intrinsic, as theoretically predicted in double-exchange systems such as manganites [33] in which a decrease of ~0.1 eV below $T_{Curie}$ could explain a part of the increase in the sampling depth at 50 K, but not account for the entire observed magnitude. Experimental studies so far have reported a change in the work function with opposite sign in double-layer LSMO [34] and in manganite grain boundaries [35]. We note that other "extrinsic" factors, such as absorption of surface contaminants upon exposure to air or during low-temperature vacuum measurements that are extended in time, could also contribute to changes of the work function from its value on a pristine surface [36].

## V. CONCLUSIONS

We have studied the TEY intensity as a function of film thickness and incidence angle in two families of perovskites, LCMO and YBCO, at the Mn and Cu $L$ edges and have extracted values for the sampling depth $d$, attenuation length $\lambda$ and their ratio $\lambda/d$. The values for $d$ found for these perovskites range between the smaller values typically found for pure metals and larger values reported for Fe oxides, and show a temperature dependence to the sampling depth in LCMO that could be investigated in other correlated oxides that exhibit metal-insulator transition phenomena. These results should encourage an investigation of sampling depths in XAS measurements of other oxides, especially in light of the complicated phase diagrams they can exhibit, in order to exploit the usefulness of XAS techniques to probe the orbital, charge and magnetic properties at surfaces and buried interfaces, and for which proper interpretation requires a quantitative knowledge of the depth sensitivity of the TEY signal.

Acknowledgments

B. A. D. and A. V. acknowledge support from the FVG Regional Project SPINOX funded by Legge Regionale 26/2005 and the decreto 2007/LAVFOR/1461. R.W. gratefully acknowledges support by the Cusanuswerk, Bischöfliche Studienförderung. Beamtime at Elettra Synchrotron (Trieste, Italy) was performed at APE beamline under proposal 20105081. A. R. and B. A. D. gratefully acknowledge Francesca Veracini of the Casalis-Scandolo Institute for familial assistance during beamtimes. This work was funded in part by the Deutsche Forschungsgemeinschaft (Project No. KO 1303/8-1).

APPENDIX A – Electron yield production and attenuation length calculations

Under X–ray irradiation, the electron yield signal is made up of two contributions of electrons escaping from the sample surface: 1) the primary electrons, constituted by a small number of excited photoelectrons and Auger electrons extracted from the atomic potential during photoabsorption and radiationless deexcitation; and 2) the secondary electrons, produced along the Auger electrons' trajectory during scattering principally with band electrons and at a lower extent with other core electrons [31]. The subsequent inelastic scattering that the secondary electrons suffer during their migration towards the free sample surface results in a broad kinetic energy-distribution function. The TEY absorption edge signal corresponds to the integration of a photoelectrons spectrum, from the high-kinetic energy valence band photoelectrons to the low-kinetic energy secondary electrons produced in the electron cascades (0–30 eV). Even considering specific variations between materials, typically the secondary electron spectrum has a full width at half-maximum below 10 eV and it is peaked below ~3–4 eV, with the majority of electrons having an energy below 5 eV and 2 eV respectively for metals and insulators. For hard X–rays the photoelectron spectrum is dominated by Auger electrons, as secondary electrons are mostly generated at a depth greater than their escape depth and cannot leave the sample. In contrast, in the soft X–ray range (~50 –1000 eV) cascades of secondary electrons are the major contribution (~80 – 90%) [29, 37]. Some authors, though, consider the component from primary Auger electrons and from high-energy secondary electrons not negligible as compared to low-energy secondaries, where the latter may be reduced due to a smaller than expected mean free path of secondary electrons or because of reflections losses at the sample surface [9, 30, 38, 39]. Only a fraction of the secondary electrons generated by internal processes in the near-surface volume escape from the sample: those that reach the surface with a kinetic energy sufficient to overcome the vacuum barrier. The average depth of emission of these electrons is defined as the escape depth $\lambda_{escape}$. For secondary electrons it is related to the inelastic mean free path (i.e. the mean distance an electron travels without any inelastic scattering event) projected along the sample surface normal and averaged for different secondary electron energies. This distance is generally much smaller (<1 nm) than the range of primary Auger electrons (5 – 50 nm) generated by soft X–rays. $R_{Auger}$ is related to the initial electron kinetic energy $E$ and to the material density $\rho$ by the approximate empirical formula $R_{Auger} \approx E^{1.4}/\rho$, in which $R_{Auger}$ is expressed in Å, $E$ in eV and $\rho$ in g/cm$^3$ [40]. This power law relationship with energy is valid in the energy range 1 – 10 keV. With a rough approximation [31] it may be applied to Auger electrons at the Mn $L_{2,3}$ edge ($L_3M_{2,3}M_{2,3}$ (544eV), $L_3M_{2,3}V$ (585 eV), $L_3VV$ (636 eV) ) obtaining $R_{Auger} \approx 8$ nm. The escape depth, and consequently the sampling depth, is highly dependent on the crystallographic and electronic structure of the specific material and it is difficult to estimate. For example, in the transition metals, the number of holes in the *d*-band is considered the most important parameter to determine *d* [41].

Calculations of the attenuation length in materials can be done using tabulated values for atomic photoabsorbtion cross-sections that do not include any absorption fine structure (or near–edge autoionization resonances i.e. white lines) [42]. In multicomponent systems the absorption cross section is the sum of individual coefficients each multiplied by the weight fraction present. For LCMO the calculated atomic photoabsorption cross section is $\mu_a^{LCMO}(E) = \{0.7\,\sigma_a^{La}(E) + 0.3\,\sigma_a^{Ca}(E) + \sigma_a^{Mn}(E) + 3\,\sigma_a^{O}(E)\}N_A\rho^{LCMO}$, where $N_A$ is Avogadro's number and $\rho^{LCMO}$ is the molar density of LCMO. We scaled the pre–$L_3$ and post–$L_2$ edge regions of Mn to the off–resonance absorption $\mu_a^{LCMO}(E)$ previously published. This allows the conversion from the arbitrary units of the yield spectra into an absolute absorption coefficient scale (units of length$^{-1}$).

REFERENCES

[1] F. de Groot and A. Kotani, Core Level Spectroscopy of Solids (CRC Press FL, 2008).

[2] J. Stöhr, NEXAFS Spectroscopy (Springer, New York, 1996).

[3] B. H. Frazer, B. Gilbert, B. R. Sonderegger and G. de Stasio, Surf. Sci. **537**, 161 (2003).

[4] G. Akgül, F. Aksoy, Y. Ufuktepe and J. Lüning, Solid State Comm. **149**, 384 (2009); G. Akgül, F. Aksoy, A. Bozduman, O. M. Ozkendir, Y. Ufuktepe and J. Lüning, Thin Sol. Films, **517**, 1000 (2008).

[5] Y. Ufuktepe, G. Akgül and J. Lüning, J. Alloys and Comp. **401** 193 (2005).

[6] R. Nakajima, J. Stöhr and Y. U. Idzerda, Phys. Rev. B **59**, 6421 (1999).

[7] V. Chakarian, Y.U Idzerda and C. T. Chen, Phys. Rev. B **57**, 5312 (1998).

[8] J. Vogel, and M. Sacchi, J. Elec. Spectr. Relat. Phenom. **67**, 181 (1994).

[9] M. Abbate, J.B. Goedkoop, F.M.F. de Groot, M. Grioni, J. C. Fuggle, S. Hofmann, H. Petersen and M. Sacchi, Surf. Interf. Anal. **18,** 65 (1992).

[10] R. G. Jones and D. P. Woodruff, Surf. Sci. **114**, 38 (1982).

[11] S. Gota, M. Gautier-Soyer, and M. Sacchi, Phys. Rev. B **62**, 4187 (2000).

[12] J. Lüning, F. Nolting, A. Scholl, H. Ohldag, J. W. Seo, J. Fompeyrine, J.-P. Locquet and J. Stöhr, Physical Review B, **67**, 214433 (2003).

[13] J. Chakhalian, J. W. Freeland, H.-U. Habermeier, G. Cristiani, G. Khaliullin, M. van Veenendaal and B. Keimer, Science **318**, 1114 (2007).

[14] R. Werner, C. Raisch, A. Ruosi, B. A. Davidson, P. Nagel, M. Merz, S. Schuppler, M. Glaser, J. Fujii, T. Chassé, R. Kleiner and D. Koelle, Phys. Rev. B **82**, 224509 (2010).

[15] C. T. Chen, Y. U. Idzerda, H.-J. Lin, N. V. Smith, G. Meigs, E. Chaban, G. H. Ho, E. Pellegrin and F. Sette, Phys. Rev. Lett. **75**, 152 (1995).

[16] B. T. Thole, P. Carra, F. Sette and G. van der Laan, Phys. Rev. Lett. **68**, 1943 (1992).

[17] C. Piamonteze, P. Miedema and M.F. de Groot, Phys. Rev. B, **80**, 184410 (2009).

[18] V. Chakarian and Y. Idzerda, J. Appl. Phys. **81**, 4709 (1997).

[19] J. Hunter Dunn, D. Arvanitis, N. Martensson, M. Tischer, F. May, M. Russo and K. Baberschke, J. Phys. Cond. Mat. **7**, 1111 (1995).

[20] J. Vogel and M. Sacchi, Phys. Rev. B **49**, 3230 (1994).

[21] W.L. O'Brien and B. P. Tonner, Phys. Rev. B **50**, 12672 (1994).

[22] Y. U. Idzerda, C. T. Chen, H.-J. Lin, G. Meigs, G. H. Ho and C.-C. Kao, Nucl. Instr. Meth. Phys. Res. A **347**, 134 (1994).

[23] B. T. Thole, G. van der Laan, J. C. Fuggle, G. A. Sawatzky, R. C. Karnatak and J.-M. Esteva, Phys. Rev. B **32**, 5107 (1985).

[24] M. B. Salamon and M. Jaime, Rev. Mod. Phys. **73**, 583 (2001); Y. Tokura, Rep. Prog. Phys. **69**, 797 (2006).

[25] G. Panaccione, I. Vobornik, J. Fujii, D. Krizmancic, E. Annese, L. Giovanelli, F. Maccherozzi, F. Salvador, A. De Luisa, D. Benedetti, A. Gruden, P. Bertoch, F. Polack, D. Cocco, G. Sostero, B. Diviacco, M. Hochstrasser, U. Maier, D. Pescia, C. H. Back, T. Greber, J. Osterwalder, M. Galaktionov, M. Sancrotti and G. Rossi, Rev. Sci. Instr. **80**, 043105 (2009).

[26] R. Herger, P. R. Willmott, C. M. Schlepütz, M. Björck, S. A. Pauli, D. Martoccia, B. D. Patterson, D. Kumah and R. Clarke, Y. Yacoby and M. Döbeli, Phys. Rev. B **77**, 085401 **(**2008**)**.

[27] J. O'Donnell, M. Onellion, M. S. Rzchowski, J. N. Eckstein and I. Bozovic, Phys. Rev. B **54**, R6841 (1996).

[28] T. Girardeau, J. Mimault, M. Jaouen, P. Chartier and G. Tourillon, Phys. Rev. B 46, 7144 (1992).

[29] B. L. Henke, J. A. Smith and D. T. Attwood, J. Appl. Phys. **48**, 1852 (1977); B. L. Henke, J. Liesegang and S. D. Smith, Phys. Rev. B **19**, 3004 (1979); B. L. Henke, J. P. Knauer and K. Premaratne, J. Appl. Phys. **52**, 1509 (1981).

[30] H. Henneken, F. Scholze and G. Ulm, J. Appl. Phys. **87**, 257 (2000).

[31] A. Erbil, G. S. Cargill, R. Frahm and R. F. Boehme, Phys. Rev. B **37**, 2450 (1988).

[32] A. J. Millis, Nature **392**, 147 (1998); J.-H. Park, E. Vescovo, H.-J. Kim, C. Kwon, R. Ramesh and T. Venkatesan, Nature **392**, 794 (1998).

[33] N. Furukawa, J. Phys. Soc. Jpn. **66**, 2523 (1997).

[34] K. Schulte, M. A. James, L. H. Tjeng, P. G. Steeneken, G. A. Sawatzky, R. Suryanarayanan, G. Dhalenne and A. Revcolevski, Phys. Rev. B **64**, 134428 (2001).

[35] J. Klein, C. Höfener, S. Uhlenbruck, L. Alff, B. Büchner and R. Gross, Europhys. Lett. **47**, 371 (1999).

[36] A. Shih, J. Yater, C. Hor, R. Abrams, Appl. Surf. Sci. **111**, 251 (1997).

[37] G. Martens, P. Rabe, G. Tolkiehn and A. Werner, Phys Stat. Sol. (a) **55**, 105 (1979).

[38] A. Krol, C.J. Sher and Y.H. Kao, Phys Rev B **42**, 3829 (1990).

[39] S. L. M. Schoeder, G. D. Moggridge, R. M. Ormerod, T. Rayment and R. M. Lambert, Surf. Sci. **324**, 371 (1995).

[40] L. Reimer and B. Volbert, Scanning **2**, 238 (1979).

[41] H. C. Siegmann, J. Phys.: Condens. Matter 4, 8395, (1992); H. Schönhense and H. C. Siegmann, Ann. Phys. **2**, 465 (1993).

[42] B. L. Henke, E.M. Gullikson and J.C. Davis, Atomic Data and Nuclear Data Tables, **54**, 181-342 (1993).

# Electron sampling depth and saturation effects in LCMO (and YBCO) films investigated by soft x-ray absorption spectroscopy

FIG. 1. (color online) Oscillations of the RHEED specular spot intensity during the growth of LCMO thin films. For each unit cell deposited, the intensity shows a cyclic oscillation of intensity, allowing the calibration of film thickness to much better than a unit cell ($c$ = 0.383 nm). RHEED intensity remains at the level of the starting STO during the entire growth. Data taken from LCMO film with final thickness of 13 uc.

FIG. 2. (color online) XRD $2\theta$-$\theta$ scan of LCMO thin films of different thicknesses on STO: 100uc (blue), 50uc (red) and 25uc (black). Inset shows a blowup of the thickness fringes near the (002) reflection.

FIG. 3. Resistivity as a function of temperature for the 50 uc LCMO film. The metal-insulator transition temperature (TMI) is near $T_{Curie} \approx 160$ K, and the residual resistivity at 4 K is ~100 $\mu\Omega$–cm.

FIG. 4. Peak heights of absorption intensity $I_{TEY}$ (difference of peak-to-preedge intensities, as discussed in the text) at the Mn $L_{2,3}$ edges plotted as a function of LCMO film thickness $t$, at room temperature. Fit of $A(1-\exp(-t/d))$ to the data gives a sampling depth $d$ = 2.6 ± 0.3 nm for $L_3$ and $d$ = 2.7 ± 0.3 nm $L_2$ edges. Data points are solid circles for the $L_3$ edge and open circles for the $L_2$ edge; fits are solid lines.

FIG. 5. Experimental absorption coefficient $\mu(E)$ of LCMO at 300 K, 150 K and 50 K (solid lines) from normalization of the absorption intensity spectra to the calculated atomic photoabsorption cross section of LCMO (dashed line) from Ref. [last Henke-34]. The curves are nearly indistinguishable from each other, indicating the temperature-independence of $\mu$ and hence of $\lambda = 1/\mu$.

FIG. 6 (color online) (a) $I_{TEY}(E)$ for different incidence angles $\theta$ across the Mn $L_{2,3}$ edge for the 50 uc thick LCMO film at 300 K, showing the effects of saturation on overall $I_{TEY}$ intensity and relative spectral $L_3$ and $L_2$ intensity. (b) $I_{TEY}\sin\theta$ for different incidence angles at the Mn $L_2$ and $L_3$ peaks at 300 K (taken from the spectra in (a)), 150 K and 50 K. Best-fit lines shown in the figures give values for $\lambda/d$ = 15.4 ($L_3$ edge) and 23.7 ($L_2$ edge) at 300 K, and 11.1 ($L_3$ edge) and 17.0 ($L_2$ edge) at 50 K.

FIG. 7 (color online) (a) $I_{TEY}(E)$ for different incidence angles $\theta$ across the Cu $L_{2,3}$ edge for the 200 nm YBCO film at 150 K, showing the effects of saturation (b) $I_{TEY}\sin\theta$ for different incidence angles at the Cu $L_2$ and $L_3$ peaks at 150 K (taken from the spectra in (a)) and 50 K. Best-fit lines shown in the figures give values for $\lambda/d$ = 19.1 ($L_3$ edge) and 25.2 ($L_2$ edge) at 150 K, and 21.3 ($L_3$ edge) and 28.3 ($L_2$ edge) at 50 K.

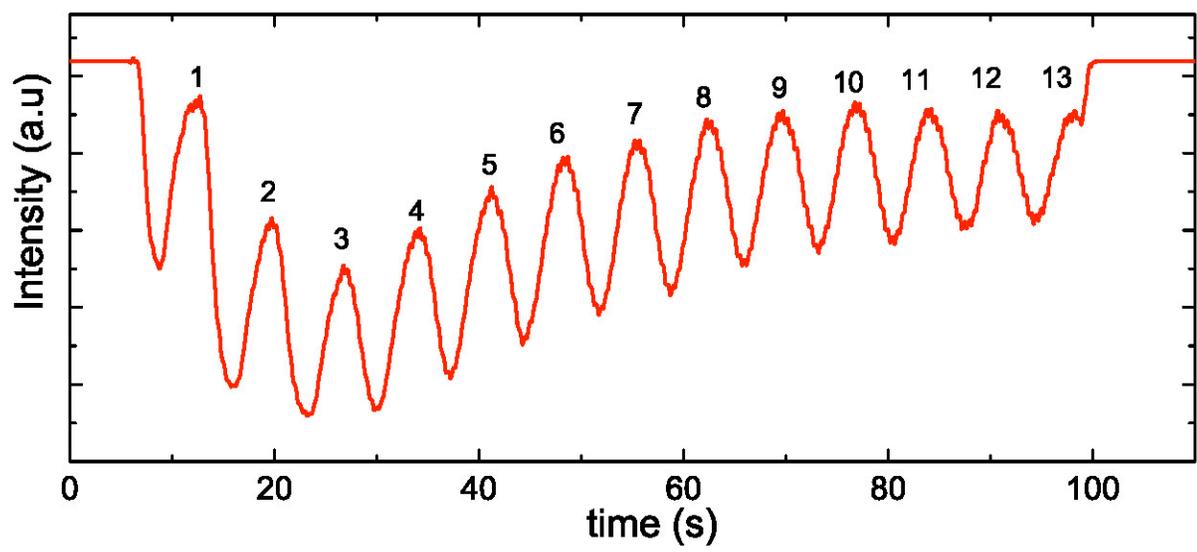

FIG. 1 – Ruosi et al.

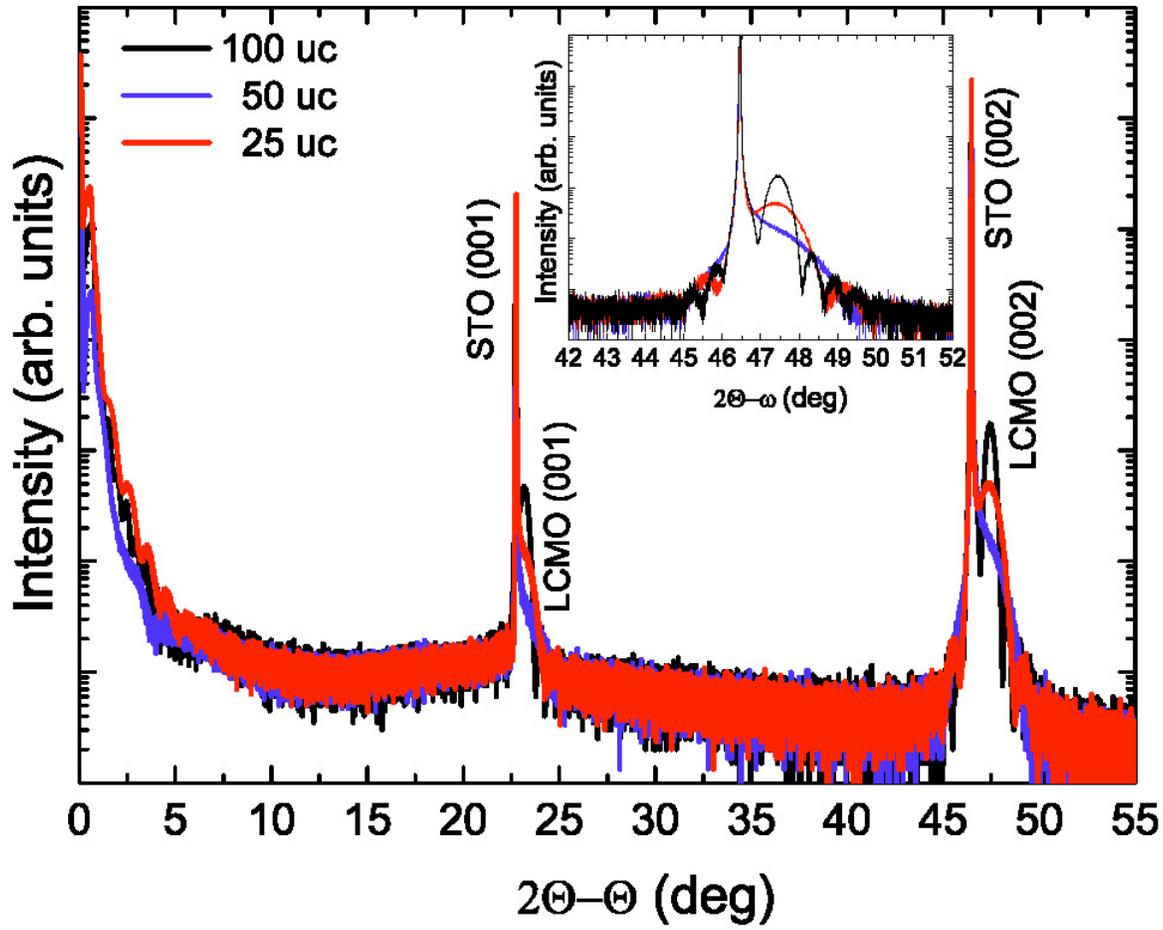

FIG. 2 – Ruosi et al.

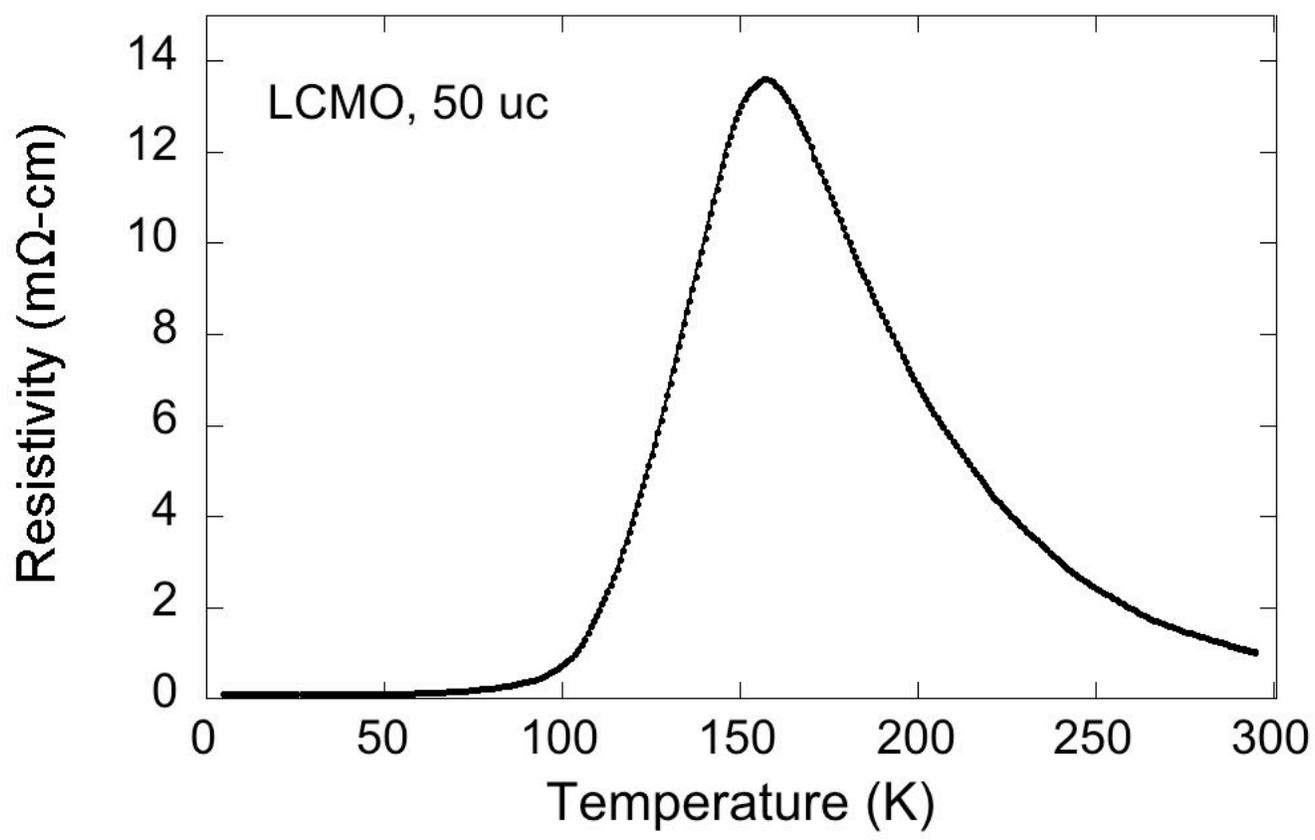

FIG. 3 – Ruosi et al.

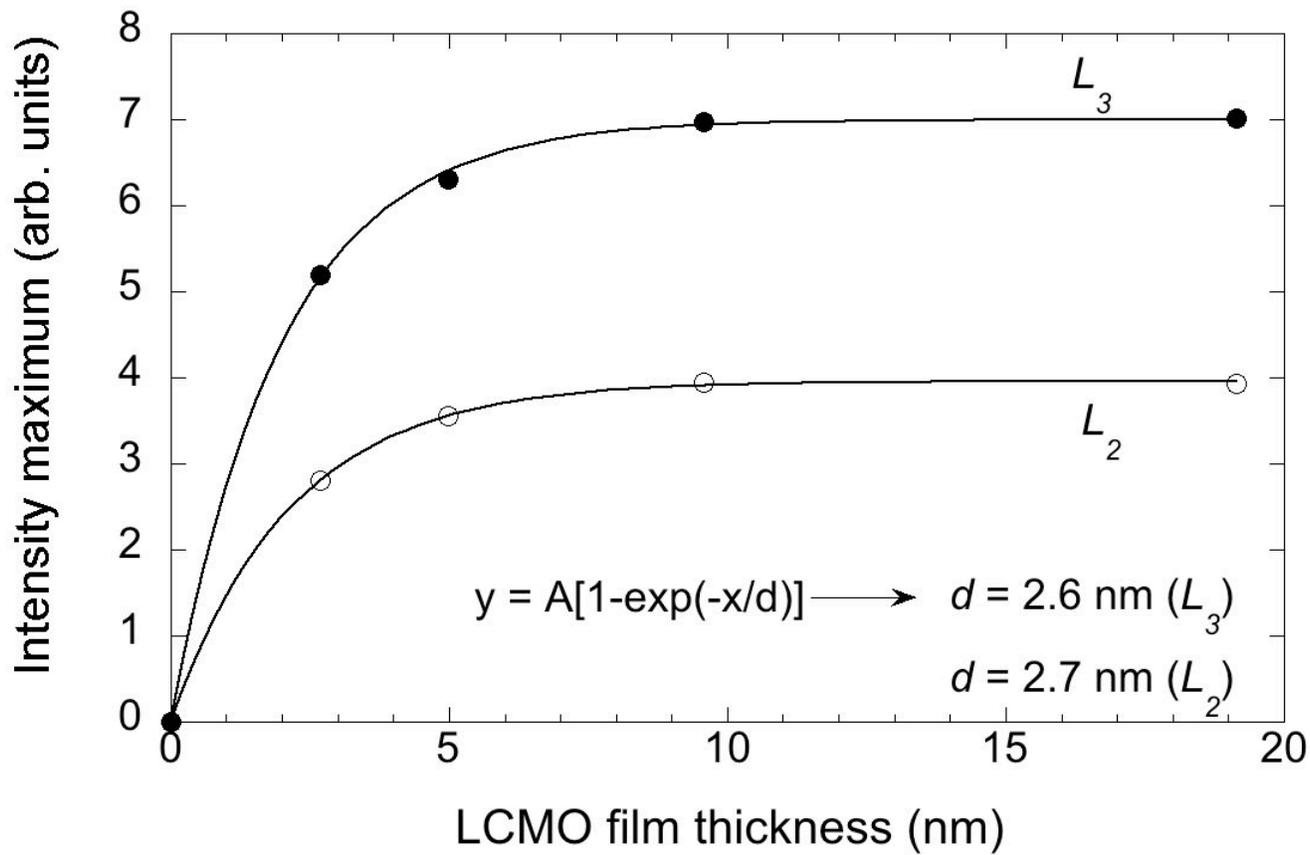

FIG. 4 – Ruosi et al.

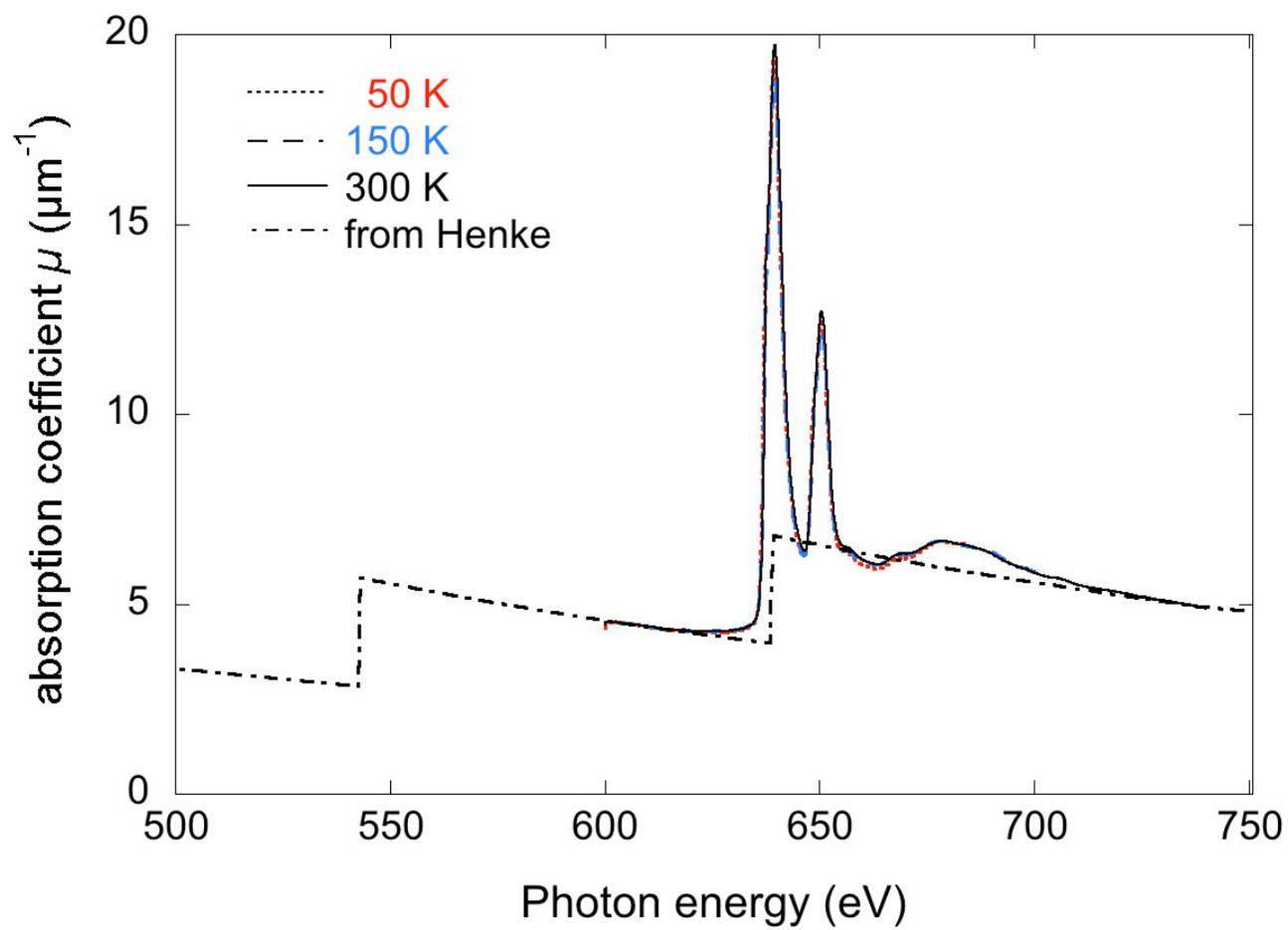

FIG. 5 – Ruosi et al.

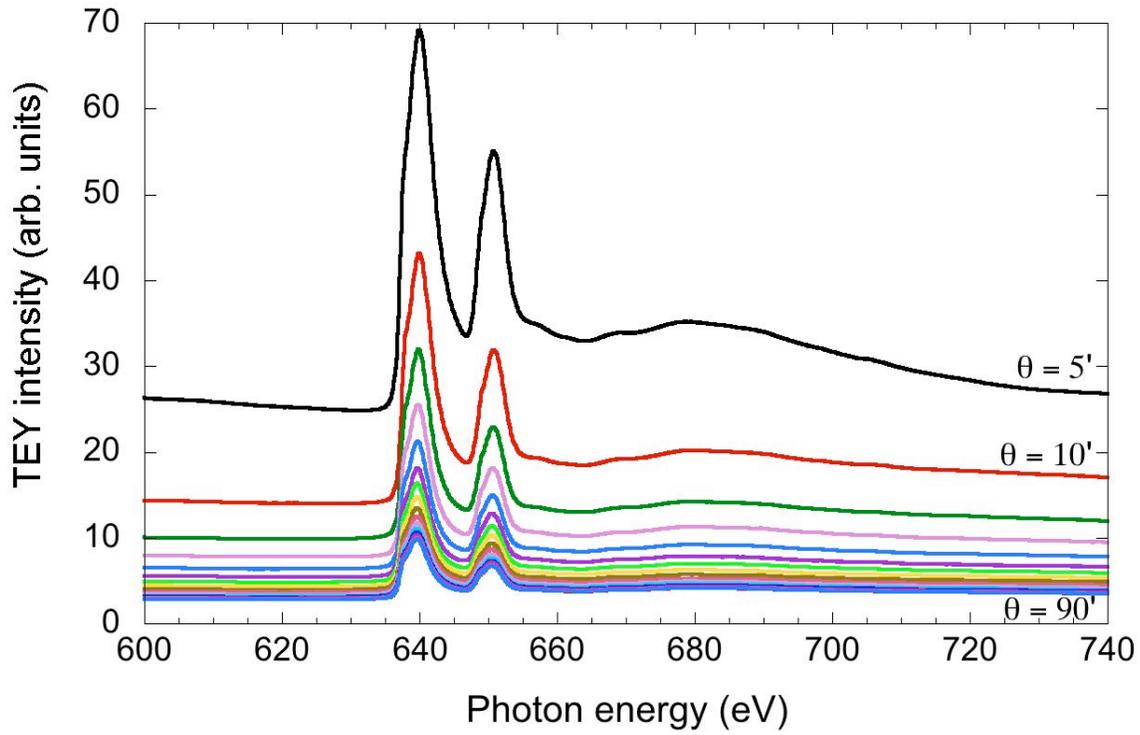

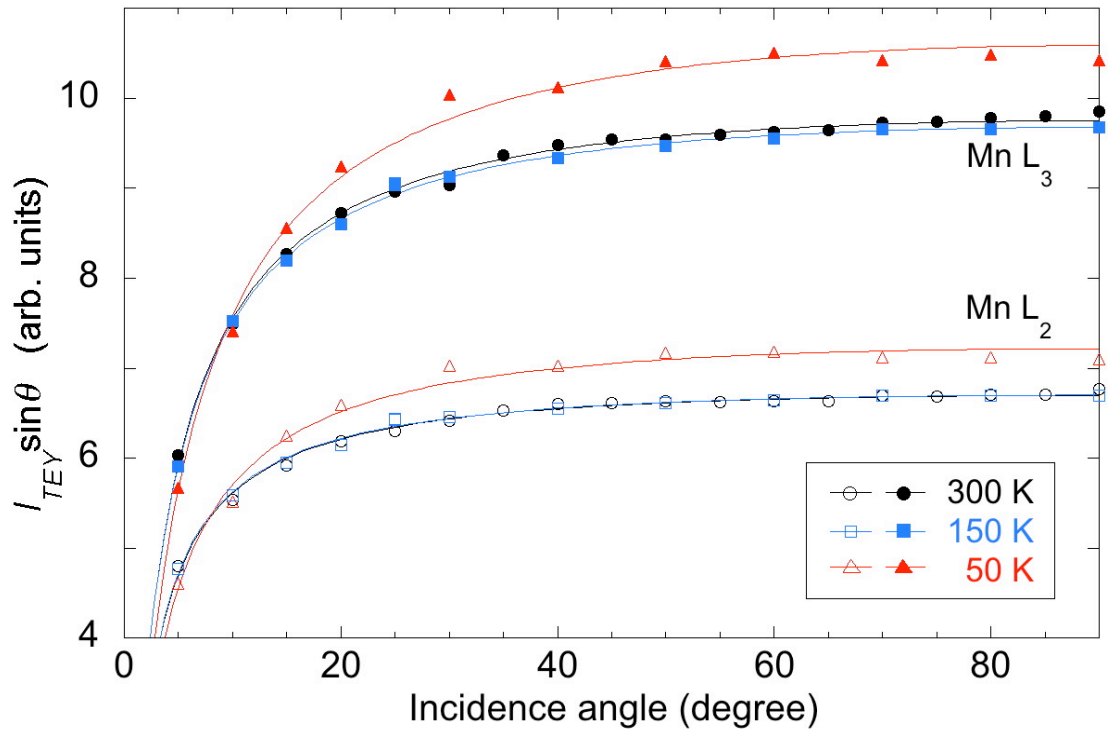

FIG. 6 – Ruosi et al.

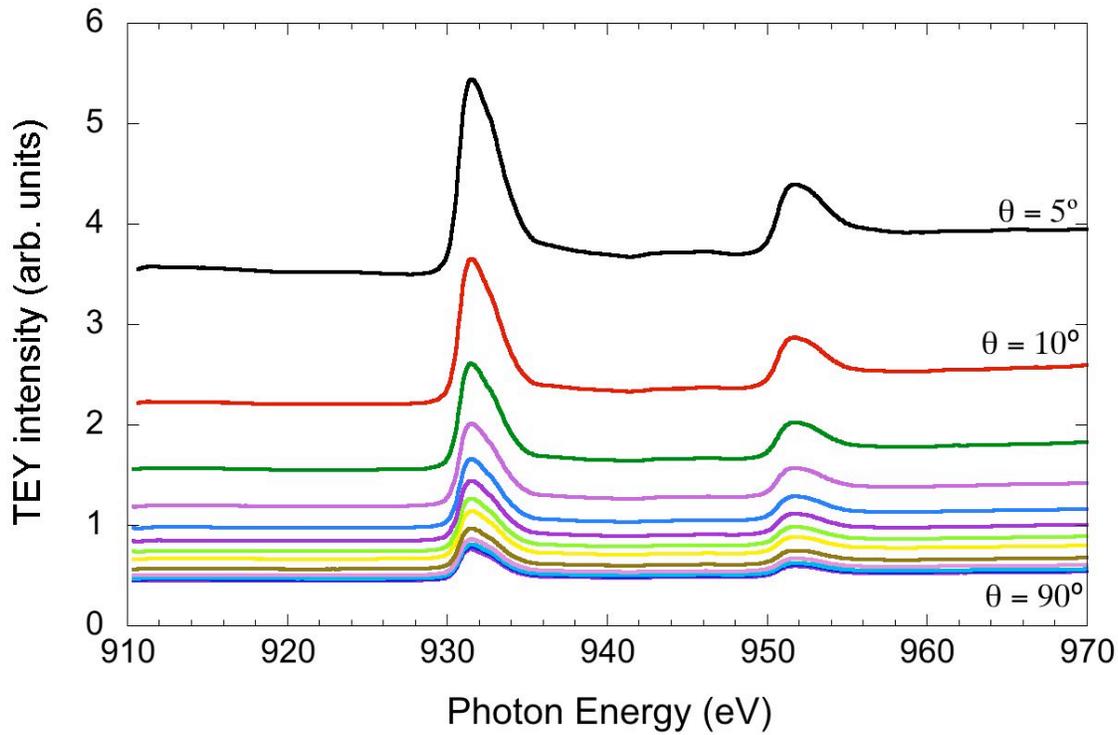
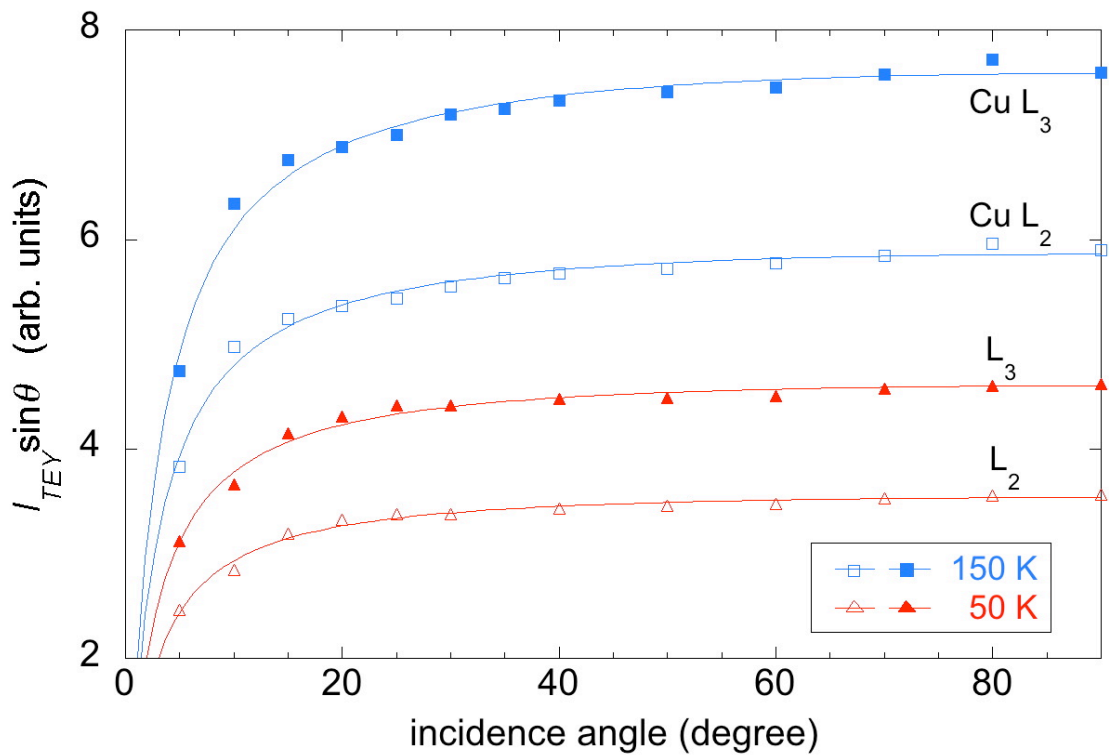

FIG. 7 – Ruosi et al.